\documentclass[12pt]{iopart}
\usepackage{graphicx}

\usepackage{iopams}  
\usepackage{braket}
\usepackage{dsfont}
\usepackage{bbold}

\begin{document}

\title[Amplification of mechanical quadratures using weak values]{Amplification of mechanical quadratures using weak values}

\author{Sergio Carrasco$^1$ and Miguel Orszag$^{1,2}$}

\address{$^1$ Instituto de F\'isica, Pontificia Universidad Cat\'olica de Chile, Casilla 306, Santiago, Chile}
\address{$^2$ Centro de \'{O}ptica e Informaci\'{o}n Cu\'{a}ntica, Camino la Pir\'{a}mide 5750, Huechuraba, Santiago, Chile}
\eads{\mailto{sjcarras@uc.cl}, \mailto{morszag@fis.puc.cl}}
\vspace{10pt}
\begin{indented}
\item[]February 2022
\end{indented}

\begin{abstract}
An interferometric arrangement is proposed in which the technique of weak value amplification is implemented in order to enlarge the effect of a single photon on the quadratures of a movable mirror of an optical cavity. The photon interacts weakly with the mirror via radiation pressure and is postselected in the dark port of the interferometer. The real and imaginary parts of weak values of angular momentum type photonic operators produce an amplification of the mirror quadratures, which is large as compared to the scenario in which all photons are taken into consideration, i.e. when no postselection is performed. The effect is studied both for a mirror initialized in a thermal and coherent states. For a thermal state, the weak value amplification effect is boosted with the number of particles of the mirror, which occurs due to the imaginary part of the weak values. 
\end{abstract}

%
\vspace{2pc}
\maketitle
%
%

\section{Introduction}

In this article we will study the interaction between a single photon (a microscopic degree of freedom) and a movable mirror of a Fabry-P\'{e}rot cavity (a macro or  mesoscopic degree of freedom). Inside the cavity, the photon and the mirror  interact via radiation pressure \cite{Law,Optomechanics1,Optomechanics2} which produces optomechanical entanglement \cite{Marshall2003}. Outside the cavity, the emitted photon is detected in the dark port of an interferometric arrangement, that selects a certain photonic pure state. This is a probabilistic procedure called \emph{postselection} \cite{PS1}, that may fail or be successful. In the latter case a final state of the photon is also specified, besides the initial state (both states are thereby known).  We will study the change of the quadratures of the mirror after the interaction, conditioning on the events for which the postselection was successful. Since the interaction between the mirror and the photon is weak, the change of the mechanical quadratures will depend on \emph{weak values} of the photonic operators that take part in the interaction. 

The weak value of an observable is a complex value that shows up in a weak measurement of the observable followed by postselection \cite{Aha,Sve,Tam,Kof,Dre}. The magnitude of the weak value can lie outside the eigenvalue range, in which case it is said to be anomalous. An anomalous weak value will have therefore a larger effect on the measurement device as compared to a standard measurement, in which case the pointer variable of the apparatus is displaced by a quantity within the eigenvalue range. The increased effect of the system on the measurement device due to anomalous weak values will be referred as the weak value amplification effect. Weak values are objects interesting on their own \cite{Matzkin,Vaidman,WeakValuesEPR}, that have been used to study counterfactual paradoxes \cite{Hardy1,Hardy2,3box1,3box2}, wave-particle duality \cite{DoubleSlit1,DoubleSlit2} and are related to contextuality \cite{Pusey}.  They have applications in quantum state reconstruction \cite{State1,State2,State3} and metrology \cite{Hall,Dixon,Starling,OAM,Hallaji,Cheshire}. In metrological applications, although the measurement precision does not improve with the use of weak values, it is possible the reach the same level of precision with fewer data (since postselection reduces the size of the data set) \cite{Jordan,Zhang,Alves,Sinclair}. Consequently, the effect of weak values in the radiation pressure interaction may be useful for the estimation of optomechanical parameters. Additionally, weak values have been considered as a quantum effect \cite{Quant1,Quant2}, which makes interesting their application in mesoscopic systems \cite{Simon}.     

The structure of this article is as follows. In section \ref{section2} we review the weak value amplification effect. Our treatment, however, considers a different interaction hamiltonian between the system and the measurement device, as compared to the standard hamiltonian presented in most literature on the subject. In section \ref{section3} we show how the previous effect may be applied to the interaction between a single photon and the moving mirror of an optical cavity. In section \ref{section4} our results are presented for a mirror initialized in a thermal and coherent states. Finally, in section \ref{section5} the results are commented and summarized. 

\section{Weak value amplification}\label{section2}

It will be assumed that the \emph{measurement device} (MD) is a quantum harmonic oscillator, while the \emph{target system} (or simply, the system) is a quantum system with total angular momentum described by the eigenvalue equation $\hat{J}^2\ket{j,m}=j(j+1)\ket{j,m}$, where $\{\ket{j,m},m=-j...j\}$ represents the angular momentum basis whose elements span any system state. The initial quantum states of the MD and the system are denoted as $\rho_{M}$ and $\rho_S$, respectively, while the initial joint state of the system and the MD is the product (uncorrelated) state $\rho_{S,M}=\rho_{S}\otimes\rho_{M}$.  Expectation values on these initial states will be denoted as $\langle \hat{M} \rangle\equiv\Tr(\hat{M}\rho_{M})$ and $\langle \hat{S} \rangle\equiv\Tr(\hat{M}\rho_{S})$, where $\hat{M}$ and $\hat{S}$ are hermitian operators of  the MD and the system, respectively. The interaction between the system and the MD is described by the hamiltonian 
\begin{eqnarray}\label{IntHamiltonian}
\hat{H}_{int}=g\delta(t)(\hat{J}_{x}\hat{Y}+\hat{J}_{y}\hat{X}),
\end{eqnarray}
where ${J}_{x}, J_{y}$ are angular momentum operators of the system, and $\hat{X},\hat{Y}$ are the quadratures of the MD, which satisfy $[\hat{X},\hat{Y}]=i$. The function $\delta(t)$ is a Dirac delta function and $g$ is a real parameter representing the coupling strength between the system and the MD. It is worth to point out that this hamiltonian differs from the interaction hamiltonian of a measurement, since the latter corresponds to the product of a system variable (the measured observable) and an apparatus variable, a model often referred as the von Neumann hamiltonian \cite{Sve,vonNeumann}.  When the free evolution of the MD and the system can be neglected, the interaction hamiltonian generates the evolution operator 
\begin{eqnarray}\label{EvolOperator}
\hat{U}=\exp{[-ig(\hat{J}_{x}\hat{Y}+\hat{J}_{y}\hat{X})]}.
\end{eqnarray}
Let $\hat{R}=\hat{S}\otimes\hat{M}$ be a joint observable of the system and the MD. The expectation value of $\hat{R}$ on the state after the interaction  is $\langle\hat{R}\rangle_f\equiv \Tr(\hat{R}\hat{U}\rho_{S,M}\hat{U}^{\dagger})=\Tr(\hat{U}^{\dagger}\hat{R}\hat{U}\rho_{S,M})$. The transformation $\hat{U}^{\dagger}\hat{R}\hat{U}$ can be expanded in powers of $g$ using the well known operator expansion \cite{Orszag}, 
\begin{eqnarray}\label{OpExpansion}\fl
\hat{U}^{\dagger}\hat{R}\hat{U}=\hat{R}+ig[\hat{J}_{x}\hat{Y}+\hat{J}_{y}\hat{X},\hat{R}]+\frac{(ig)^2}{2!}[\hat{J}_{x}\hat{Y}+\hat{J}_{y}\hat{X},[\hat{J}_{x}\hat{Y}+\hat{J}_{y}\hat{X},\hat{R}]]+...
\end{eqnarray}
If the interaction between the MD and the system is sufficiently weak so that the terms up to second order are small as compared with the first two terms of the expansion, then 
\begin{eqnarray}\nonumber
\langle\hat{R}\rangle_f\approx \langle\hat{S}\rangle\langle\hat{M}\rangle&+&i(g/2)\langle [\hat{J}_x,\hat{S}]\rangle\langle\{\hat{Y},\hat{M}\}\rangle+i(g/2)\langle \{\hat{J}_x,\hat{S}\}\rangle\langle[\hat{Y},\hat{M}]\rangle\\ \label{WeakMeas}
&+&i(g/2)\langle [\hat{J}_y,\hat{S}]\rangle\langle\{\hat{X},\hat{M}\}\rangle+i(g/2)\langle \{\hat{J}_y,\hat{S}\}\rangle\langle[\hat{X},\hat{M}]\rangle,
\end{eqnarray}
where $[\cdot,\cdot]$ and $\{\cdot,\cdot\}$  are commutators and anticommutators, respectively.   We will distinguish now two measurements strategies: a) without postselection and b) with postselection, which will be described below. 

\subsection{Measurement strategy without postselection}

In this case $\hat{R}=\mathbb{1}\otimes\hat{M}$, which means that the variable $\hat{M}$ of the MD is observed in order to extract some information regarding to the system. Expression (\ref{WeakMeas}) reduces to
\begin{eqnarray}\label{WithoutPS}
\langle\hat{M}\rangle_f\approx\langle\hat{M}\rangle+ig\langle\hat{J}_x\rangle\langle[\hat{Y},\hat{M}]\rangle+ig\langle\hat{J}_y\rangle\langle[\hat{X},\hat{M}]\rangle,
\end{eqnarray}
which shows that the change of the $X$ quadrature of the MD will be $g\langle\hat{J}_x\rangle$, while the shift of the $Y$ quadrature will be $-g\langle\hat{J}_y\rangle$. In general, the change of $\hat{M}$ will be proportional to the expectation values of $\hat{J}_x$ and $\hat{J}_y$. This is what is expected from a weak measurement without postselection: the shift of the pointer will depend linearly on the average initial value of the system variables that appear in the interaction hamiltonian (\ref{IntHamiltonian}). 

\subsection{Measurement strategy with postselection}

In this scenario we consider first the case $\hat{R}=\hat{P}_{\phi}\otimes\mathbb{1}$, where $\hat{P}_{\phi}=\ket{\phi}\bra{\phi}$ is a projector into the pure (normalized) system state $\ket{\phi}$. In this situation, expression (\ref{WeakMeas}) becomes
\begin{eqnarray}\label{PostselectionProbability}
\langle\hat{P}_{\phi}\rangle_f\approx\langle\hat{P}_{\phi}\rangle+ig\langle [\hat{J}_x,\hat{P}_{\phi}]\rangle\langle \hat{Y} \rangle+ig\langle [\hat{J}_y,\hat{P}_{\phi}]\rangle\langle \hat{X} \rangle.
\end{eqnarray} 
This result corresponds to the probability to read the outcome ``$1$'' in a strong measurement of the projector, i.e. if the system and the MD interact weakly via the hamiltonian (\ref{IntHamiltonian}) and then we measure strongly the projector $\hat{P}_{\phi}$,  the probability to read the eigenvalue $1$ will be given by this expression. This probability is also called the probability to postselect the state $\ket{\phi}$ after the weak interaction between the MD and the system. Notice that, when the expectation values of the commutators vanish, then the probability simply corresponds to $\langle \hat{P}_{\phi}\rangle$, i.e. to the probability to project the initial state into the final state $\ket{\phi}$. Note also that the approximation (\ref{PostselectionProbability}) assumes that $\langle \hat{P}_{\phi}\rangle$ is larger than the higher order terms of the expansion, and thereby can not be arbitrarily small. 

Suppose now that we perform a measurement of $\hat{R}=\hat{P}_{\phi}\otimes\hat{M}$ and ask for the expectation value of $\hat{M}$ conditioned to the cases in which the system state $\ket{\phi}$ was successfully postselected. The conditioned expectation value $\mathbb{E}(\hat{M}|f)$ corresponds to the ratio 
$\langle{\hat{P}_{\phi}\hat{M}}\rangle_f / \langle \hat{P}_{\phi}\rangle_f$, where the numerator can be evaluated using (\ref{WeakMeas}) while the denominator corresponds to (\ref{PostselectionProbability}). The final result is
\begin{eqnarray}\nonumber
\mathbb{E}(\hat{M}|f)=\langle \hat{M}\rangle&+& ig\frac{ \langle \{ \hat{J}_x,\hat{P}_{\phi}\} \rangle }{2\langle\hat{P}_{\phi}\rangle }\cdot  \langle [\hat{Y},\hat{M}]\rangle-g\frac{ \langle [ \hat{J}_x,\hat{P}_{\phi}]\rangle }{i\langle\hat{P}_{\phi}\rangle }\cdot Cov(\hat{Y},\hat{M})\\  \label{WithPS}
&+&ig\frac{ \langle \{ \hat{J}_y,\hat{P}_{\phi}\} \rangle }{2\langle\hat{P}_{\phi}\rangle }\cdot \langle [\hat{X},\hat{M}]\rangle-g\frac{ \langle [ \hat{J}_y,\hat{P}_{\phi}]\rangle }{i\langle\hat{P}_{\phi}\rangle }\cdot Cov(\hat{X},\hat{M}),
\end{eqnarray}
where $Cov(\hat{A},\hat{B})=\langle \{\hat{A},\hat{B}\} \rangle/2-\langle \hat{A}\rangle \langle \hat{B}\rangle$ is the covariance between the variables $\hat{A}$ and $\hat{B}$. This expression constitutes a direct application of equation (17) of \cite{Jozsa2007} to a hamiltonian of the type (\ref{IntHamiltonian}). It shows that the change of $\hat{M}$ depends both on the commutators and anticommutators between the system variables ($\hat{J}_x,\hat{J}_y$ and $\hat{P}_{\phi}$), which are normalized by $\langle \hat{P}_{\phi} \rangle$. Therefore, the change may be  larger than the one described by (\ref{WithoutPS}). When the initial system state is pure, i.e. $\rho_S=\ket{\psi}\bra{\psi}$, these terms are related to the real and imaginary parts of \emph{weak values}, 
\begin{eqnarray}
\Re(J_{x,w})=\frac{ \langle \{ \hat{J}_x,\hat{P}_{\phi}\} \rangle }{2\langle\hat{P}_{\phi}\rangle },\quad
\Im(J_{x,w})=-\frac{ \langle [ \hat{J}_x,\hat{P}_{\phi}]\rangle }{2i\langle\hat{P}_{\phi}\rangle },\\ \label{WeakValues1}
\Re(J_{y,w})=\frac{ \langle \{ \hat{J}_y,\hat{P}_{\phi}\} \rangle }{2\langle\hat{P}_{\phi}\rangle },\quad
\Im(J_{y,w})=-\frac{ \langle [ \hat{J}_y,\hat{P}_{\phi}]\rangle }{2i\langle\hat{P}_{\phi}\rangle },
\end{eqnarray}
where $\Re(\star)$ and $\Im(\star)$ denote the real and imaginary parts of  $\star$, while the complex values $J_{x,w}$ and $J_{y,w}$ correspond to the weak values of $\hat{J}_{x}$ and $\hat{J}_y$, respectively, which are defined as
\begin{eqnarray}\label{WeakValues2}
J_{x,w}=\frac{ \bra{\phi} \hat{J}_x \ket{\psi} } {\langle \phi | \psi \rangle}, \quad
J_{y,w}=\frac{ \bra{\phi} \hat{J}_y \ket{\psi} } {\langle \phi | \psi \rangle}.
\end{eqnarray}
The shift of $\hat{M}$ depends therefore on both weak values, which may be larger than the expectation values appearing in (\ref{WithoutPS}). In order to amplify the change of $\hat{M}$ it is useful to chose quasi orthogonal initial and final states. However, there is a limit to the amplification with weak values: the zero and first order terms of (\ref{WeakMeas}) and (\ref{PostselectionProbability}) should dominate over the higher order terms of the expansion. 

\section{Application to optomechanical system}\label{section3}

\begin{figure}[h!]
\begin{center}
\includegraphics[scale=0.42]{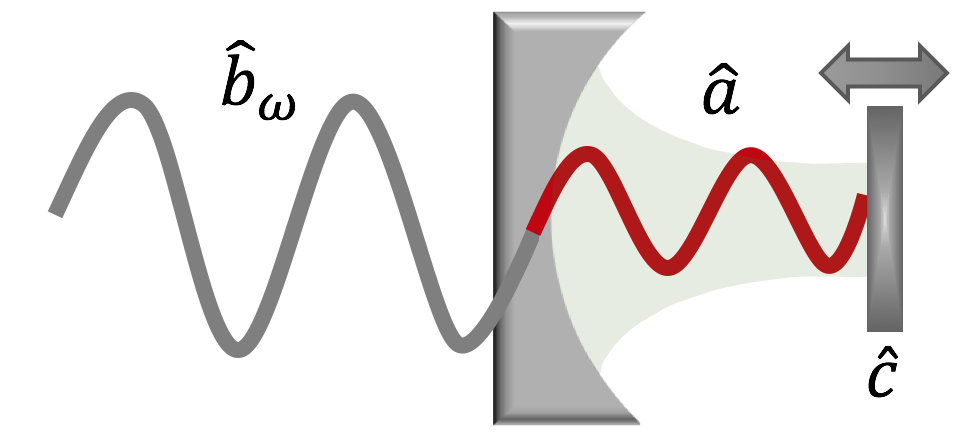}
\caption{A single cavity mode interacts with a single mechanical mode (the center of mass of the movable mirror) via radiation pressure. The cavity mode in turn interacts with a set of infinite (sinusoidal) external modes. }
\label{figure1}
\end{center}
\end{figure}

In this section we show how the previous results can be applied to an optomechanical system (OMS). The OMS corresponds to an optical Fabry P\'{e}rot cavity with one movable mirror. The center of mass of the mirror is described as a single mode quantum harmonic oscillator of frequency $\Omega$. The cavity mode is coupled to an infinite collection of external modes, as it is depicted in figure \ref{figure1}. The hamiltonian $\hat{H}$ that describes the dynamics of the whole system (the OMS coupled to the external modes) is given by
\begin{eqnarray}\label{OMSExtHamiltonian} \nonumber
\hat{H}&=&\hat{H}_{OMS}+\hat{H}_{f}+\hat{H}_{dis},\\ \nonumber
\hat{H}_{OMS}&=&\omega_{cav}\hat{a}^{\dagger}\hat{a}+\Omega\hat{c}^{\dagger}\hat{c}-g_0\hat{a}^{\dagger}\hat{a}(\hat{c}^{\dagger}+\hat{c}),\\ \nonumber
\hat{H}_{f}&=&
\int_0^{\infty}d\omega
\omega\hat{b}^{\dagger}_{\omega}\hat{b}_{\omega},\\
\hat{H}_{dis}&=&\sqrt{\frac{\Gamma}{2\pi}}\Big(\hat{a}\int_0^{\infty}d\omega\hat{b}^{\dagger}_{\omega}+\hat{a}^{\dagger}\int_0^{\infty}d\omega\hat{b}_{\omega}\Big).
\end{eqnarray}
The first contribution, $\hat{H}_{OMS}$, describes the OMS. The cavity and mechanical mode operators are $\hat{a}$ and $\hat{c}$, respectively. The frequency of the cavity mode is denoted as $\omega_{cav}$. The parameter $g_0$ is the vacuum optomechanical coupling strength, that quantifies the strength of the radiation pressure interaction between a single cavity mode and the moving mirror. The second contribution, $\hat{H}_f$, is the free energy of the external modes. The third term, $\hat{H}_{dis}$, describes the coupling between the external field and the cavity mode,  where $\Gamma$ is the decay rate of the cavity field through the fixed mirror. 

We will assume that the initial state of the external field is a (pure) single photon state,\begin{eqnarray}\label{initialsystemstate}
\ket{\psi}_{S}=\int_{0}^{\infty}d\omega G(\omega;\omega_0,\epsilon)\hat{b}_{\omega}^{\dagger}\ket{\emptyset}_S, \quad G(\omega;\omega_0,\epsilon)=\sqrt{\frac{\epsilon}{\pi}}\frac{1}{\omega-\omega_0+i\epsilon},
\end{eqnarray}
where $|G(\omega;\omega_0,\epsilon)|^2$ is a Lorentzian spectral distribution. The parameter $\omega_0$ is the median of the distribution and $\epsilon$  is the half-width at half-maximum of the spectrum (a measure of the width of the distribution). The state $\ket{\emptyset}_S$ is the multimode vacuum state.  

If the moving mirror starts in a number state $\ket{n}_{M}$ and the cavity is initially empty $\ket{0}_{cav}$, the Schr\"{o}dinger equation can be solved analytically.  In an interaction picture with respect to $\Omega\hat{c}^{\dagger}\hat{c}+\hat{H}_{f}$, the evolution of the external field and the OMS corresponds to
\begin{eqnarray}\label{Process1}
\ket{n}_{M} \ket{\psi}_{S}\rightarrow\ket{n}_{M} \ket{\psi}_{S}-\sqrt{2}g\hat{J}_{+}\ket{\psi}_S\hat{c}\ket{n}_{M}+\sqrt{2}g\hat{J}_{-}\ket{\psi}_S\hat{c}^{\dagger}\ket{n}_{M}.
\end{eqnarray}
The operators $\hat{J}_{+},\hat{J}_{-}$ are ladder (angular momentum type)  operators that will create single photons states with a shifted frequency spectrum. We will describe them below, but previously let us comment on different aspects regarding the process described by equation (\ref{Process1}). In the first place, note that the probabilities to excite the oscillator to the state $\ket{n+1}_{M}$, or to produce a transition to $\ket{n-1}_{M}$, are proportional to $g^2$, where $g\equiv g_0/\Omega$ is a small scaled optomechanical parameter. Therefore, the initial states will be barely perturbed. Secondly, notice that cavity states do not appear in expression (\ref{Process1}). This occurs because the cavity begins and ends in the vacuum state, and therefore it its not necessary to include it explicitly. In the third place, and as it is explained in detail in \ref{A}, the process described by (\ref{Process1}) is valid under different conditions, which are; \textbf{a)} the OMS should operate in the resolved side band regime ($\Omega\gg\Gamma$), \textbf{b)} the radiation pressure interaction should be weak ($g^2n\ll1$), \textbf{c)} the photon should have a small detuning with respect to the cavity frequency, i.e. $\omega_0=\omega_{cav}+\delta$, where the detuning is $\delta\equiv-g_0^2/\Omega$, and \textbf{d)} be nearly monochromatic ($\epsilon\ll\Gamma$), and \textbf{e)} the interaction should last for long times ($t\gg\epsilon^{-1}$). Now, let us turn into the definition of $\hat{J}_{+},\hat{J}_{-}$,  considering the following boson creation operators
\begin{eqnarray}
\eqalign{
\hat{b}^{\dagger}_{-1}\equiv\int_{0}^{\infty}d\omega G(\omega;\omega_0-\Omega,\epsilon)\hat{b}_{\omega}^{\dagger},\qquad
\hat{b}^{\dagger}_{0}\equiv\int_{0}^{\infty}d\omega G(\omega;\omega_0,\epsilon)\hat{b}_{\omega}^{\dagger},\\
\hat{b}^{\dagger}_{1}\equiv\int_{0}^{\infty}d\omega G(\omega;\omega_0+\Omega,\epsilon)\hat{b}_{\omega}^{\dagger}.}
\end{eqnarray}
It is easy to note that $\ket{\psi}_S=\hat{b}_{0}^{\dagger}\ket{\emptyset}_S$, and $\ket{d}_S\equiv\hat{b}_{-1}^{\dagger}\ket{\emptyset}_S$ is a single photon state with the frequency spectrum ``shifted to the left'' by $\Omega$, while $\ket{u}_S\equiv\hat{b}_{1}^{\dagger}\ket{\emptyset}_S$ is a single photon state with the frequency spectrum ``shifted to the right'' by $\Omega$. Since there is no frequency overlap between the states, it is clear that $\langle \psi |_S |u \rangle_S=\langle \psi |_S | d \rangle_S=\langle d |_S |u \rangle_S=0$. Also, all the operators satisfy the usual boson commutation relations, and all the states ($\ket{\psi}_S$, $\ket{u}_S$ and $\ket{d}_S$) are normalized. Using these definitions, we introduce the ladder operators as follows, 
\begin{eqnarray}
\hat{J}_{+}\equiv\sqrt{2}(\hat{b}_{1}^{\dagger}\hat{b}_{0}+\hat{b}_{0}^{\dagger}\hat{b}_{-1}),\qquad\hat{J}_{-}=\hat{J}_{+}^{\dagger}.
\end{eqnarray}
Notice that $\hat{J}_{+}\ket{\psi}_S=\sqrt{2}\ket{u}_S$ and $\hat{J}_{-}\ket{\psi}_S=\sqrt{2}\ket{d}_S$, i.e. the ladder operators applied to the initial state simply shift its spectrum to the left or to the right. Making an analogy to a system with angular momentum, the state $\ket{\psi}_S$ would be a state with angular momentum numbers $j=1,m=0$, the state $\ket{u}_S$ would be defined by $j=1,m=1$ and the state $\ket{d}_S$ by $j=1,m=-1$. 

Thereby, when the initial state of the external field is $\ket{\psi}_S$, the cavity is empty, and conditions \textbf{a}-\textbf{e} are satisfied, then the interaction between the external field and the moving mirror, can be described by an effective evolution operator  
\begin{eqnarray}\label{EffectiveEvolOp}
\hat{U}=\exp{[\sqrt{2}g(\hat{J}_{-}\hat{c}^{\dagger}-\hat{J}_{+}\hat{c})]}=\exp{[-ig(\hat{J}_{x}\hat{Y}+\hat{J}_{y}\hat{X})]},
\end{eqnarray}
where $\hat{J}_{x},\hat{J}_{y}$ are defined from $\hat{J}_{\pm}=\hat{J}_{x}\pm i\hat{J}_{y}$, while the mirror quadratures are $\hat{X}=(\hat{c}^{\dagger}+\hat{c})/\sqrt{2}$ and $\hat{Y}=i(\hat{c}^{\dagger}-\hat{c})/\sqrt{2}$. In the right hand side of the second identity, the parameter $g$ has been redefined as $g\equiv 2g$ (to avoid carrying the factor of 2).  Consequently, the results presented in the previous section can be applied to the OMS coupled to an external field, and both measurement strategies (with and without postselection) can be compared. 

\section{Results}\label{section4}

Let us consider first the measurement without postselection, which is described by equation (\ref{WithoutPS}). In order to account for the free evolution of the mirror, the variable $\hat{M}$ of the right hand side should be replaced by $\hat{M}(t)=\exp{(i\Omega\hat{c}^{\dagger}\hat{c}t)}\hat{M}\exp{(-i\Omega\hat{c}^{\dagger}\hat{c}t)}$. Note that given the initial system state (\ref{initialsystemstate}), the expectation values $\langle \hat{J}_x\rangle=\langle \hat{J}_y\rangle=0$. Therefore, there is no change of $\hat{M}$ to first order in $g$ besides its free evolution, i.e. $\langle \hat{M}\rangle_f=\langle\hat{M}(t)\rangle$. However, by using a different initial state $\langle \hat{J}_x\rangle\sim\langle \hat{J}_y\rangle\sim1$, i.e. the expectation values may lie in the eigenvalue range of the variables. Consider the observation of the $X$ and $Y$ quadratures of a mirror initially prepared in a thermal or coherent state. In these cases, expression (\ref{WithoutPS}) becomes
\begin{eqnarray}\label{XYthermalCoherentNPS}
\fl
\langle \hat{X} \rangle_f=\langle\hat{X}(t)\rangle+g[\cos(\Omega t)-\sin(\Omega t)],\quad
\langle \hat{Y} \rangle_f=\langle\hat{Y}(t)\rangle-g[\sin(\Omega t)+\cos(\Omega t)],
\end{eqnarray}
which shows that both variables oscillate with an amplitude equal to the coupling constant $g$. Recall that the terms $\langle\hat{X}(t)\rangle$ and $\langle\hat{Y}(t)\rangle$ correspond to the free evolution of the quadratures. For a thermal state both mean values are zero, while for a coherent state $\ket{\alpha}$ the expectation values are $\langle\hat{X}(t)\rangle=\sqrt{2}[\cos(\Omega t)\Re(\alpha)+\sin(\Omega t)\Im(\alpha)]$ and $\langle\hat{Y}(t)\rangle=\sqrt{2}[\cos(\Omega t)\Im(\alpha)-\sin(\Omega t)\Re(\alpha)]$.

\begin{figure}[h!]
\begin{center}
\includegraphics[scale=0.42]{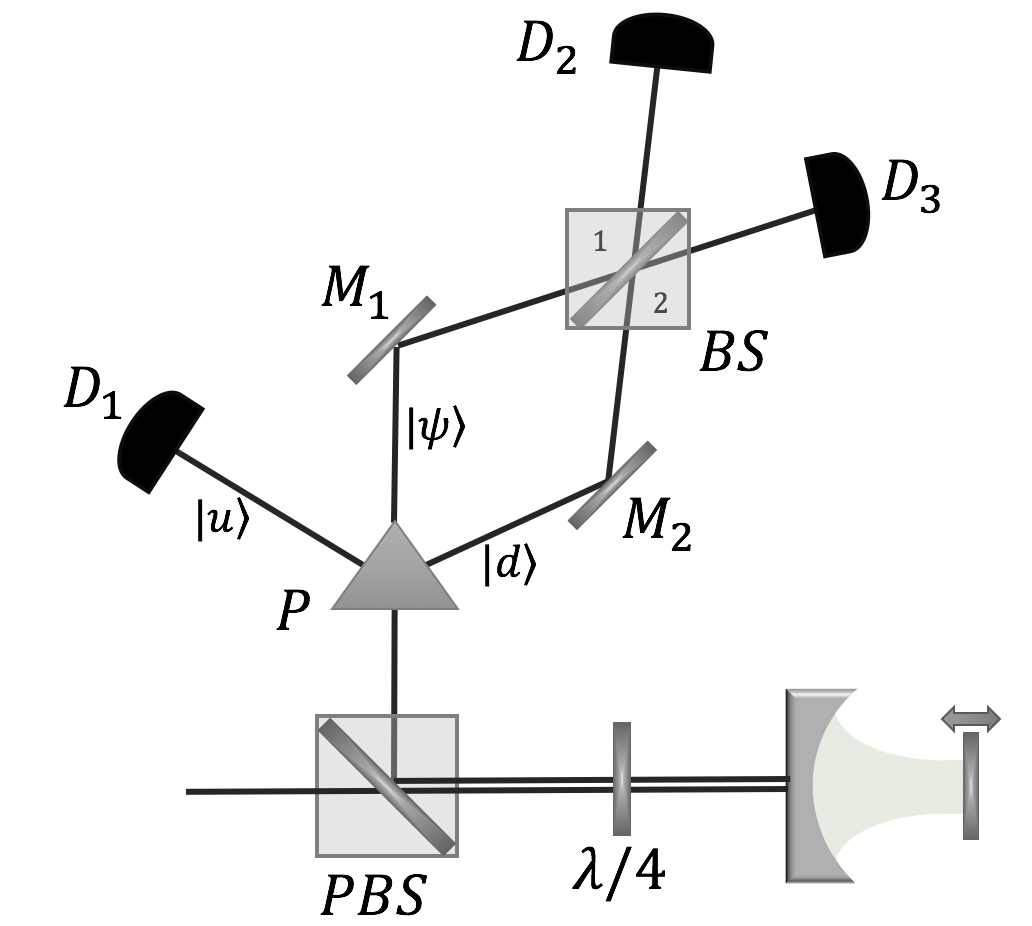}
\caption{Interferometric arrangement proposed to select the state (\ref{PSstate}) by photon detection in the dark port (detector $D3$). The light exiting from the OMS is split by a prism ($P$) into different paths, two of which interfere in the beam splitter (BS). The set of reflectances ($r_1,r_2$) and transmittances ($t_1,t_2$) of BS are defined as $r_1=\sqrt{1-\delta^2}e^{-i(\theta-\pi/2)},t_1=\delta e^{-i\theta},r_2=\sqrt{1-\delta^2}e^{i\pi/2},t_2=\delta$.}
\label{figure2}
\end{center}
\end{figure}

On the other hand, in order to consider the strategy with postselection a method to select a pure state should be implemented. We will consider the interferometric arrangement depicted in figure \ref{figure2}, by which the detection of a photon in the so called dark port will select the state  
\begin{eqnarray}\label{PSstate}
\ket{\phi}_S=\delta e^{i\theta}\ket{\psi}_S-i\sqrt{1-\delta^2}\ket{d}_S,
\end{eqnarray}
where $\delta$ is a real parameter between 0 and 1. According to expression (\ref{WeakValues2}), the quantum weak values of $\hat{J}_x$ and $\hat{J}_y$ are
\begin{eqnarray}
J_{x,w}=\frac{\sqrt{1-\delta^2}}{\sqrt{2}\delta}e^{i(\theta+\pi/2)},\quad J_{y,w}=\frac{\sqrt{1-\delta^2}}{\sqrt{2}\delta}e^{i(\theta+\pi)}.
\end{eqnarray}
Note that both weak values are complex values with the same amplitude but with a phase difference of $\pi/2$ (using a state $\ket{\phi}_S$ of a more general form, weak values with different amplitudes and phases can be obtained). 

Consider again the observation of the $X$ and $Y$ quadratures of a mirror  prepared in a thermal state.  Expression (\ref{WithPS}) shows that
\begin{eqnarray}\label{XthermalPS}
\mathbb{E}(\hat{X}|f)&=&2(1+N) \Big(\frac{g}{\delta}\Big)\sqrt{\frac{1-\delta^2}{2}}\sin(\Omega t-\theta), \\ \label{YthermalPS}
\mathbb{E}(\hat{Y}|f)&=&2(1+N) \Big(\frac{g}{\delta}\Big)\sqrt{\frac{1-\delta^2}{2}}\cos(\Omega t-\theta).
\end{eqnarray}
The amplification is therefore proportional to the amplitude of the weak value and the mean number of phonons $N\equiv\langle\hat{c}^{\dagger}\hat{c}\rangle$. It is worth noting that the amplification depends mainly on the \emph{imaginary part of the weak values}, which couple to the covariance terms, as may be seen in equation (\ref{WithPS}).  For a thermal state, the covariances depend on the mean number of phonons (the order of magnitude of the commutators, on the other hand, is equal to unity). 

\begin{figure}[h!]
\begin{center}
\includegraphics[scale=0.55]{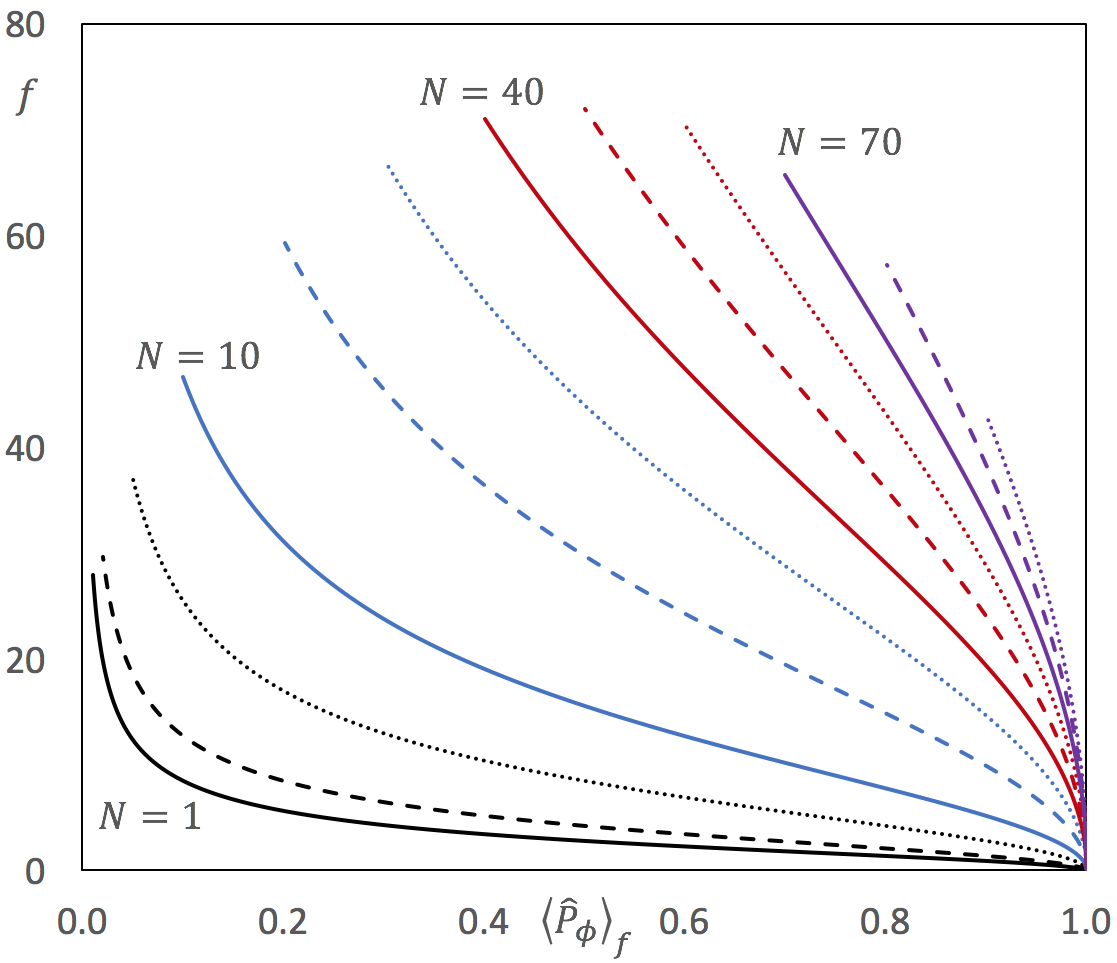}
\caption{An amplification factor $f$, defined as the amplitude of the oscillations (\ref{XthermalPS}) and (\ref{YthermalPS}), normalized by the coupling parameter $g$, is plotted as a function of the postselection probability (\ref{PostselectionProbability}), for a MD in a thermal state at different temperatures; $N=1$ (black), $N=2$ (black dashed), $N=5$ (black dotted), $N=10$ (blue), $N=20$ (blue dashed), $N=30$ (blue dotted),
$N=40$ (red), $N=50$ (red dashed), $N=60$ (red dotted),
$N=70$ (magenta), $N=80$ (magenta dashed), $N=90$ (magenta dotted). For all cases,  $\Omega=1$ MHz and $g_0=5\cdot10^2$ Hz. The domain of each curve is limited by the restriction $\delta\gg\sqrt{N}g$. Thereby, each curve was constructed considering $\delta\geq\delta_{min}\equiv 10^2g\sqrt{N}$.   }
\label{figure3}
\end{center}
\end{figure}

\begin{figure}[h!]
\begin{center}
\includegraphics[scale=0.52]{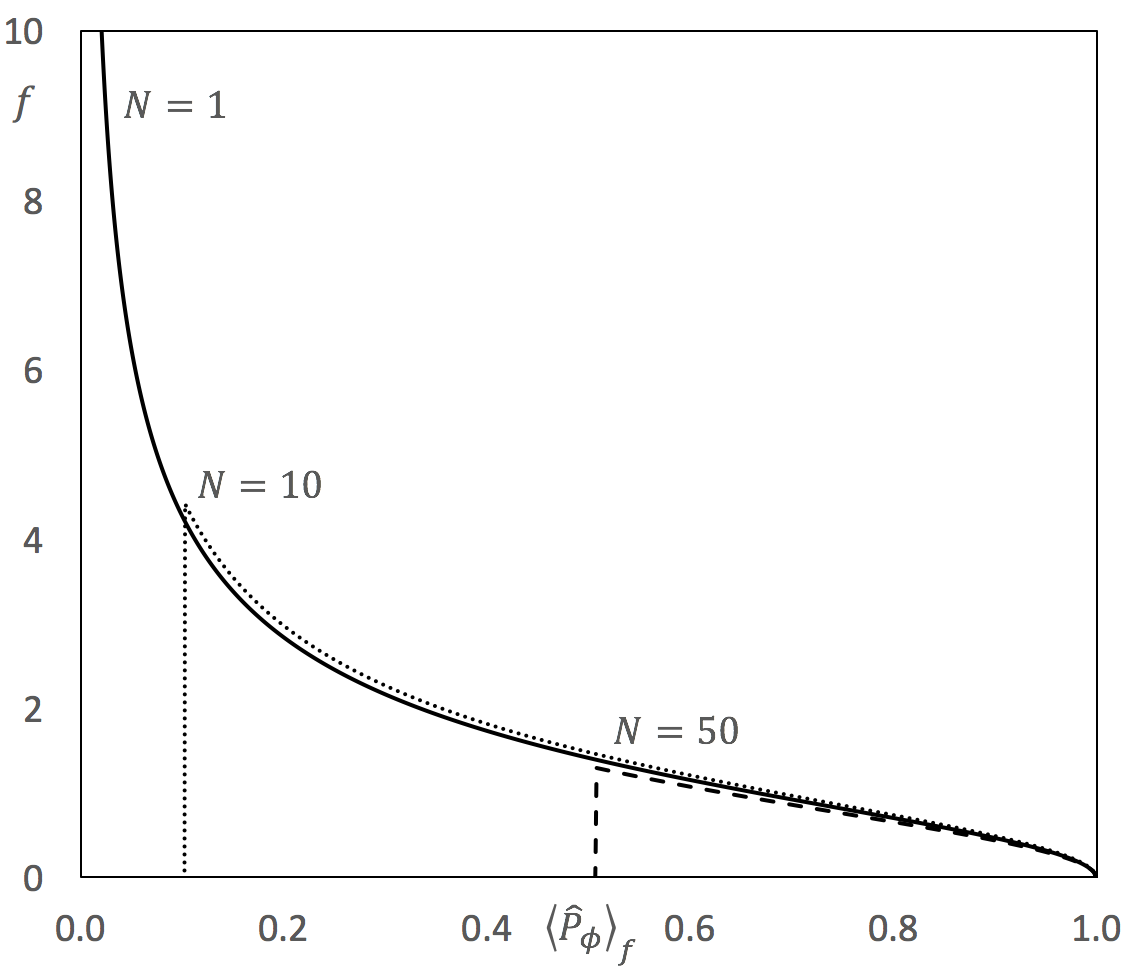}
\caption{The amplification factor $f$, defined in this case as the amplitude of the oscillations (\ref{XcoherentPS}) and (\ref{YcoherentPS}), normalized by the coupling parameter $g$, is plotted as a function of the postselection probability (\ref{PostselectionProbability}), for a MD in a coherent state with different number of phonons; $N=1$ (black), $N=10$ (black dotted) and $N=50$ (black dashed). In all cases,  $\Omega=1$ MHz and $g_0=5\cdot10^2$ Hz. As in figure \ref{figure3}, each curve was constructed considering $\delta\geq\delta_{min}\equiv 10^2g\sqrt{N}$. The postselection probability was obtained from (\ref{PSprobCoherent}), considering $\beta=\theta+\pi/2$. }
\label{figure4}
\end{center}
\end{figure}

The limit on the amplification is given by the validity of the approximations (\ref{WeakMeas}) and (\ref{PostselectionProbability}). As shown in \ref{B}, the limit  is given by $\delta\gg g\sqrt{N}$, while the postselection probability (\ref{PostselectionProbability}) corresponds to $\langle \hat{P}_{\phi} \rangle_f\approx\delta^2$. Consequently, as compared to the set of equations (\ref{XYthermalCoherentNPS}), the change of the quadratures described by  (\ref{XthermalPS}) and (\ref{YthermalPS}) may be considerable larger, as can be seen in figure \ref{figure3}. On the other hand, from equation (\ref{WithPS}) it is easy to note that for a coherent state 
\begin{eqnarray} \label{XcoherentPS}
\mathbb{E}(\hat{X}|f)&=&\langle\hat{X}(t)\rangle+2\Big(\frac{g}{\delta}\Big)\sqrt{\frac{1-\delta^2}{2}}\sin(\Omega t-\theta),\\  \label{YcoherentPS}
\mathbb{E}(\hat{Y}|f)&=&\langle\hat{Y}(t)\rangle+2\Big(\frac{g}{\delta}\Big)\sqrt{\frac{1-\delta^2}{2}}\cos(\Omega t-\theta).
\end{eqnarray}
Notice that the amplification now depends only on the magnitude of the weak values because the commutators and covariances are both of the order of the unity (unlike the previous case of a thermal state, for which the covariance depends on the number of particles). Since the limit on the amplification is defined by the restriction $g\sqrt{N}\ll \delta$, the amplification may be larger when $N$ is close to unity (the MD contains a small number of phonons), as shown in figure \ref{figure4}. 

\section{Conclusions}\label{section5}

In this work we have studied the amplification of the quadratures of a moving mirror that weakly interacts with a single photon, mediated by a Fabry-P\'{e}rot cavity, when postselection of photons is performed using an interferometric arrangement. The amplification of the effect of a single photon on the mirror quadratures depends on the weak values of two hermitian operators ($\hat{J}_x$ and $\hat{J}_y$), whose magnitudes can lie outside the eigenvalue range. 
The magnitude of the weak values is $\sim \delta^{-1}$,  where $\delta$ is a small parameter that corresponds to the transmittance of a beam splitter located at the exit of the interferometric arrangement, while the postselection probability ($\sim\delta^2$) is the fraction of light arriving at the dark port of the interferometer. The limit on the amplification by weak values is given by the restriction $ g\sqrt{N}\ll \delta \leq1$, where $g$ is a scaled coupling constant between the photon and the movable mirror of the cavity, and $N$ is the mean number of particles of the mirror.  As compared to a standard strategy that does not postselect photons (but employs all of them) the amplification gain may be large. On the other hand, the phases of the weak values may be set in order to adjust the time at which the amplification reaches its maximum value.  The real parts of the weak values couples to the commutators between the mechanical quadratures, while the imaginary parts to the covariances between the quadratures. For some states, such as thermal states, the latter depends on the mean number of particles of the MD, which increases by $N$ the weak value amplification effect produced by the single photon. Therefore, in these cases, the amplification occurs mainly due to the imaginary part of the weak values.  For  coherent states, on the contrary, both terms (commutators and covariances) are of the order of the unity and the amplification effect is based only on the anomalous weak values.  The proposed setup might be useful for the estimation of  $g$, a parameter that can be small for different optomechanical systems. 

\appendix
\section{Interaction between OMS and external modes}\label{A}
The hamiltonian (\ref{OMSExtHamiltonian}) preserves the total number of photons $\hat{a}^{\dagger}\hat{a}+\int d\omega \hat{b}^{\dagger}_{\omega}\hat{b}_{\omega}$. Since the initial state has one photon, the state of the whole system (the external optical modes, the cavity mode and the mechanical mode) at any time, $\ket{\psi(t)}$, will belong to the single photon Hilbert space, and can be expressed as
\begin{eqnarray}\fl \label{singlephotonstate}
\ket{\psi(t)}=
\ket{1}_{cav}\ket{\emptyset}_S\sum_{m=0}^{\infty}A_m(t)\ket{\tilde{m}}_M
+\ket{0}_{cav}\sum_{m=0}^{\infty}\ket{m}_M\int_0^{\infty} d\omega B_m(\omega,t)\hat{b}_{\omega}^{\dagger}\ket{\emptyset}_S,
\end{eqnarray}
where $\ket{\tilde{m}}_M$ is a \emph{displaced number state} \cite{Nieto}, defined as $\ket{\tilde{m}}_M\equiv\exp{[(g_0/\Omega)(\hat{c}^{\dagger}-\hat{c})]}\ket{m}_M$, while $A_m(t)$ and $B_m(\omega,t)$ are the probability amplitudes that satisfy the normalization condition of the state. The first term describes the situation in which the photon is inside the cavity, the oscillator in some linear combination of displaced states, and the external field in the vacuum state. The second term corresponds to the case in which the photon is outside the cavity and eventually entangled to the oscillator (while the cavity is empty). Given the initial conditions $A_m(0)=0$ and $B_m(\omega,0)=\delta_{m,n}G(\omega;\omega_0,\epsilon)$, the Schr\"odinger equation can be solved analytically. This is done in \cite{Liao}, from which we take equation (15) to point out that, for long times ($t\Gamma\gg1$  and $t\epsilon\gg1$), the solution (in a rotating frame with respect to the free hamiltonians of the mirror and the external field) is  $A_m(t)=0$ (no photons remain in the cavity) and
\begin{eqnarray}\fl
B_{m}(t)=G(\omega;\omega_0,\epsilon)\delta_{n,m}-
i\sqrt{2\pi\Gamma}
G(\omega;\omega_0+[n-m]\Omega,\epsilon)
\sum_{k=0}^{\infty}C_{m}(k)G(\omega;\Delta_{k,m},\Gamma/2),
\end{eqnarray}
where $\Delta_{k,m}=\omega_{cav}+\Omega(k-m)-g_0^2/\Omega$ and  $C_m(k)=\langle m |_M |\tilde{k}\rangle_M  \langle \tilde{k}|_M|n\rangle_M$. The Lorentzian functions $G(\omega;\Delta_{k,m},\Gamma/2)$ that appear in the summation do not overlap in the resolved sideband regime, $\Gamma\ll\Omega$. If $\omega_0$  is chosen to be $\omega_{cav}-g_0^2/\Omega$ and the spectral width of the pulse is not larger than the cavity decay rate, $\epsilon\leq\Gamma$, then all terms in the summation, for which $k\neq n$, will have a neglectable overlap with  the fist factor, $G(\omega;\omega_0+[n-m]\Omega,\epsilon)$. Therefore, in this scenario only the term $k=n$ will survive, and we can thereby perform the approximation 
 \begin{eqnarray}\fl
B_{m}(t)=G(\omega;\omega_0,\epsilon)\delta_{n,m}-
i\sqrt{2\pi\Gamma}C_{m}(n)
G(\omega;\omega_0+[n-m]\Omega,\epsilon)G(\omega;\Delta_{n,m},\Gamma/2).
\end{eqnarray}
Furthermore, if we assume that the pulse is quasi monochromatic, i.e. $\epsilon\ll\Gamma$, then the product of the two Lorentzian functions in the right hand side can be approximated by $G(\omega;\omega_0+[n-m]\Omega,\epsilon)
G(\omega=\Delta_{n,m};\Delta_{n,m};\Gamma/2)$, i.e. the second factor is replaced by its maximum value. Consequently,
\begin{eqnarray}\nonumber
B_{m}(t)=G(\omega;\omega_0,\epsilon)\delta_{n,m}-
2C_{m}(n)
G(\omega;\omega_0+[n-m]\Omega,\epsilon).
\end{eqnarray}
When $g\sqrt{n}\ll1$, then $C_n(n)=1$, $C_{n-1}(n)=-g\sqrt{n}$ and $C_{n+1}=g\sqrt{n+1}$, while the rest of the coefficients are neglectable. Therefore, $B_n(t)=-G(\omega;\omega_0,\epsilon)$, $B_{n-1}(t)=2g\sqrt{n}G(\omega;\omega_0+\Omega,\epsilon)$, and $B_{n+1}(t)=-2g\sqrt{n+1}G(\omega;\omega_0-\Omega,\epsilon)$, from which the transition (\ref{Process1}) can be directly derived by using (\ref{singlephotonstate}).

\section{Weak value amplification regime}\label{B}
In section \ref{section3} it was pointed out that  $g\sqrt{N}\ll1$ allows a first order expansion of the operator (\ref{EffectiveEvolOp}). This fact will limit the expansion (\ref{OpExpansion}) to second order terms in $g$. Therefore,
\begin{eqnarray}\fl \label{2ndOrder}
\langle{M}\rangle_f=\delta^2c_0-g\delta\sqrt{\frac{1-\delta^2}{2}}c_1+g^2\Big(\frac{1-\delta^2}{2}\Big)c_2,
\end{eqnarray}
where the coefficients $c_0$, $c_1$ and $c_2$ depend on the state of the MD. In general,
\begin{eqnarray}\fl \nonumber
c_0&=&\langle \hat{M}(t)\rangle,\\ \fl \nonumber
c_1&=& \langle\{\hat{X},\hat{M}(t)\}\rangle\sin(\theta)+ i\langle[\hat{X},\hat{M}(t)]\rangle\cos(\theta)-\langle\{\hat{Y},\hat{M}(t)\}\rangle\cos(\theta)+ i\langle[\hat{Y},\hat{M}(t)]\rangle\sin(\theta),\\ \fl
c_2&=&\langle \hat{X}\hat{M}(t)\hat{X}\rangle+
\langle \hat{Y}\hat{M}(t)\hat{Y}\rangle+
2\Im(\langle \hat{X}\hat{M}(t)\hat{Y}\rangle).
\end{eqnarray}
In order to reduce expression (\ref{2ndOrder}) to the first order expansion (\ref{WeakMeas}), we have to take into account the state of the MD. For a thermal state (and $\hat{M}$ being any quadrature), it is easy to show that $c_0=c_2=0$. Hence, expression (\ref{2ndOrder}) is automatically of first order.  For a coherent state,  two conditions will be required: \textbf{a)} $\delta^2c_0\gg g^2c_2$, and \textbf{b)} $g\delta c_1 \gg g^2c_2$.  Working out the coefficients it is possible to show that  $c_0\sim\sqrt{N}$, $c_1\sim(1+N)$ and $c_2=\sqrt{N}(N+1)$. Hence,  both conditions will be satisfied when $g\sqrt{N}\ll\delta$ (assuming  $N\geq1$). Next, let us take into consideration the postselection probability. Adding second order terms to (\ref{PostselectionProbability}),
\begin{eqnarray}\fl \label{PostselectionProbability2nOrder}
\langle \hat{P}_{\phi} \rangle_f=\delta^2-2g\delta\sqrt{\frac{1-\delta^2}{2}}\Big[\langle \hat{X}\rangle \sin(\theta)-\langle \hat{Y}\rangle \cos(\theta)\Big]+2g^2\Big(\frac{1-\delta^2}{2}\Big)(N+1).
\end{eqnarray}
For a thermal state, the first order term vanishes. Therefore, in order to disregard the second order term we need to assume that $\delta^2\gg g^2N$ (in the scenario $N\geq1$). With this assumption, the postselection probability simply corresponds to $\langle \hat{P}_{\phi} \rangle_f\approx\delta^2$. For a coherent state, on the contrary, the first order term does not vanish. There will be, as in the previous analysis, two conditions (the zero and first order terms should dominate over the the second order term). In the case $N\geq1$ both conditions are satisfied when $g\sqrt{N}\ll\delta$. In general, for a MD in a coherent state $\sqrt{N}e^{i\beta}$ the postselection probability (\ref{PostselectionProbability2nOrder}) will depend linearly on $g$, i.e. 
\begin{eqnarray}\label{PSprobCoherent}
\langle \hat{P}_{\phi} \rangle_f\approx\delta^2-2g\delta\sqrt{1-\delta^2}\sqrt{N}\sin(\theta-\beta).
\end{eqnarray} 
Consequently, and as a summary, the limit on the amplification is defined by the restriction $g\sqrt{N}\ll\delta$, both for a thermal and coherent states.


\end{document}